\documentstyle[draft,aps,prl,epsf]{revtex}
\begin{document}
\draft
\title{Orderly spectra from random interactions}

\author{C.W. Johnson$^{1}$, G.F. Bertsch$^2$, and D.J. Dean$^{3}$ }

\address{
$^1$Department of Physics and Astronomy\\ 
Louisiana State University, 
Baton Rouge, LA 70803-4001\\
$^2$Department of Physics, FM-15,  University of Washington,
Seattle, WA 98195    USA\\
$^3$Physics Division, Oak Ridge National Laboratory, P.O. Box 2008\\
Oak Ridge, Tennessee 37831 USA and Department of Physics and \\
Astronomy, University of Tennessee, Knoxville, Tennessee, 37996 
}

\maketitle

\begin{abstract}
We  investigate the low-lying spectra of many-body systems with random 
two-body interactions, specifying 
that the ensemble be   invariant under particle-hole conjugation.
Surprisingly we find patterns reminiscent of more orderly 
interactions, 
such as a predominance of 
$J=0$ ground states separated by a gap 
from the excited states, and evidence of phonon vibrations in the 
low-lying spectra.  
\end{abstract}
\copyright{1998 American Physical Society}


In the spectra of molecules, atomic nuclei, and other many-body systems,
the low-lying excitations often display a pattern 
suggestive of group symmetries, such as rotational 
or vibrational bands,  even though the many-body 
spectrum is in principle complex and the interactions themselves have 
no trace of the symmetry groups displayed.  This raises the question, 
to what extent does the low-lying spectrum acquire order simply from 
the most basic properties of the Hamiltonian?  These properties 
include rotational invariance, possibly other symmetries such as 
isospin, and the fundamental nature of the interaction, which is 
predominantly two-body in character.  Given an ensemble of 
Hamiltonians 
of this form, some properties might occur 
often, while others would occur rarely and would depend 
sensitively upon the 
detailed form of the two-body interactions.  
An example might be a 
rotational spectrum: one could imagine that a typical ground 
state might behave as a solid. 
Then many members of the ensemble would have 
a rotational band built on the ground state. 
Stated another way, many-body calculations often rely upon model 
interactions, such as pseudopotentials in atomic and molecular physics, 
and the Skyrme, quadrupole-quadrupole, and other interactions in nuclear 
physics, that despite being drastic simplifications reproduce many key 
properties.  We ask the logical extension: what properties 
remain as the Hamiltonian gets more and more arbitrary?  

In this letter, we begin exploratory studies of these questions, 
choosing ensembles of two-body random Hamiltonians 
and computing their many-body spectra. 
Although our own reference point is nuclear physics, we believe these 
issues may be relevant to generic many-body systems, such as molecules, 
atomic clusters, etc., and so our explorations should be considered 
in a broad an arena as possible. 
 Obviously the 
choice of ensemble is crucial.  In 
standard random matrix theory \cite{Mehta}, 
a powerful principle for specifying the 
ensemble is to require that it be invariant under a change of basis. 
We shall use this principle at the level of the two-body Hamiltonian to 
construct our ensembles. We first choose a single-particle basis 
labeled by 
angular momentum $j$ and  two-particle states of good total 
angular momentum $J = [ j \otimes j^\prime ]$ .
States of the same angular momentum can be transformed 
into each other, so the ensemble is specified by the average of the matrix 
elements 
and their fluctuations for each $J$.  For the 
symmetric 
matrix ensemble, invariant under orthogonal transformations, the 
mean 
square variance in matrix elements $V_{\alpha, \alpha^\prime}$ is
\begin{eqnarray}
\left \langle  
V_{\alpha, \alpha^\prime}^2 \right \rangle = 
c_J(1+\delta_{\alpha\alpha^\prime}), \\
\left \langle  V_{\alpha, \alpha^\prime} V_{\beta, 
\beta^\prime}\right \rangle 
= 0, \, \,  ( \alpha, \alpha^\prime ) \neq (\beta, \beta^\prime) 
\nonumber.
\end{eqnarray} 
Here $\alpha$ and $\alpha^\prime$ label two-body states 
$\left | j \otimes j^\prime = J \right \rangle$. Note that 
the variance depends only on $J$, and that there is 
the usual factor of 2 difference between diagonal and 
off-diagonal matrix elements.  The dependence of $c_J$ on $J$ will 
be relevant to determining the overall behavior of the ensemble.  
Obviously, 
pairing properties will dominate if $J=0$ is enhanced.  If the 
interaction 
is converted to a particle-hole representation, mean-field 
physics will 
become dominant if the diagonal $J=0$ interaction is enhanced in 
that 
representation.  For our study here, we follow the idea that the 
physics 
should be that of interacting quasiparticles, favoring neither a 
particle-particle nor a particle-hole representation.  We 
therefore demand 
that the ensemble be invariant under the Pandya transformation \cite{ph}, 
$$
\left \langle ij^{-1}; L \left | V \right | k l^{-1}; L \right 
\rangle 
= \sum_J (2J+1) \left \{ 
\matrix{  j_i & j_l & J \cr 
j_k & j_j & L } \right \} 
\left \langle il; J \left | V \right | k j; J \right \rangle .
$$
The ensemble (1) is invariant under this transformation if and 
only if 
\begin{equation}
c_J = { \bar{v}^2 \over 2J+1 }.
\end{equation}
Here $\bar{v}$ sets the energy scale for the ensemble, and our 
all results will be quoted in units of $\bar{v}$. 
Eq. (1) and (2) define the ensemble to be studied in this letter, 
which 
we term the {\it random quasiparticle ensemble} (RQE).

Random matrices were introduced into nuclear physics by Wigner \cite{Wigner} to 
model   statistical properties of nuclear spectra.
In  particular the Gaussian Orthogonal Ensemble (GOE) of random 
Hamiltonians 
describes well the level repulsion found in 
distribution of nearest-neighbor spacings of states with the same 
quantum numbers. For more global properties, however, the GOE 
does not 
match real nuclei. 
The GOE gives a semicircle level density, while realistic 
shell-model Hamiltonians tend to give a Gaussian level 
density.
But a GOE corresponds to Hamiltonians with interactions of all 
possible 
particle ranks, whereas shell-model Hamiltonians are only 
two-body 
interactions.  Wong and French \cite{Wong} investigated the 
{\it  two-body random ensemble}, 
or TBRE (also sometimes termed the {\it embedded GOE} or EGOE), 
which is similar to our RQE except that $c_J =$ constant.  
With the TBRE one regains Gaussian level densities and
Mon and French \cite{Mon} related the global level density
 to the moments of the ensemble.
All these studies, however, only considered states with 
identical quantum numbers. In contrast, our work here examines 
the relation between states of different quantum numbers.

We stress that our Hamiltonians drawn from the RQE have no symmetries 
imposed on them beyond that of Eqs.~(1), (2) above.  This is in 
contrast to earlier work \cite{Edwards76,Cortes82} which studied linear 
combinations of a random Hamiltonian and a Hamiltonian containing a 
specified symmetry (e.g. $SU(3)$ in~\cite{Cortes82}). 
These papers investigated  the relative strength of 
the random Hamiltonian necessary  to 
 overwhelm the externally imposed symmetry.  Rather than 
considering the interplay of a specified symmetry and a 
random Hamiltonian, we look to see what symmetries, or at least what 
markers of symmetries, can arise spontaneously in generic Hamiltonians.

We computed the low-lying spectra of random Hamiltonians  
for several different shell-model spaces.  
We label our systems by $N$, the number of identical particles, and 
$\Omega$, the number of single-particle states.  
For the latter we consider two different single particle spaces, first 
a space with 
$j$-orbitals  $\left\{ {1 \over2}, {3\over 2}, {5\over 2} \right \} $, 
with $\Omega = 12$, and
also in a space with $j$-orbitals  
$\left\{ {1 \over2}, {3\over 2}, {5\over 2}, {7 \over 2}\right \}$, 
with $\Omega = 20$.  In nuclear physics these correspond to the 
$1s_{1/2}$-$0d_{3/2}$-$0d_{5/2}$  and 
$1p_{1/2}$-$1p_{3/2}$-$0f_{5/2}$-$0f_{7/2}$ spaces, respectively. 
We considered $N=6$ identical particles for both the $\Omega = 12$ and 
$20$ spaces.  A nuclear spectroscopist would identify these as 
$^{22}$O and $^{46}$Ca, respectively, but because our Hamiltonians have 
been significantly abstracted we prefer the abstract labeling scheme of 
$N=6, \Omega = 12$ and $N=6, \Omega=20$.

In nuclear physics there is along with angular momentum an additional 
symmetry, isospin (which we remind our non-nuclear readers is an $SU(2)$ 
symmetry between neutrons and protons and which is a nearly exact symmetry of 
the strong nuclear force).  Since neutron-proton correlations might  allow 
different statistical behavior, we enlarge the RQE to include isospin $T$ 
which is treated exactly as $J$ in our previous definition, 
so that $c_{J,T} = \bar{v}^2 /(2J+1) (2T+1)$. For two-body interactions 
only the $T=0,1$ channels are possible.  We studied the system 
with four protons and four neutrons (and thence  $T_z = 0$) in the 
$\Omega=12 $ space, which 
corresponds to $^{24}$Mg, but which we label as $N=4, Z=4, \Omega=12$.

Before giving our results, we review the generic  
phenomenological features of the low-lying spectra of even $N$, even $Z$ 
nuclides. In Nature, all even-even nuclei have $J=0$ ground states 
which are pushed 
down in energy relative to the ground states of even-odd and odd-odd nuclei. 
The low-lying spectra display marked regularities, particularly in the 
spacing of the first $J=0,2,4,6,8, \ldots$ states. 
One labels such regularities as `vibrational' or `rotational' bands 
depending if the excitation energy goes like $J$ or $J(J+1)$, respectively. 
Other regularities are also observed and associated with various 
group structures \cite{group}, 
but these are the most basic feature. 

With this in mind, 
we now group our results under several major headings.
We computed 1000 spectra for each system, with the Hamiltonians  
drawn from the RQE as defined in Eqn.~1,2.  
 All single-particle energies were set to zero.

{\bf Predominance of $J=0$ ground states}: 
For all our ensembles we found a predominance of $J=0$  
ground states.  This is listed in Table I as a percentage. 
For the case with isospin, $N=4,Z=4, \Omega=12$, we also required 
that the ground state have $T=0$. (The other two cases with 
six identical particles automatically 
have $T=T_z=3$.)  We see that between two-thirds 
and three-quarters of the spectra have the singlet state as the lowest.
This is not a trivial consequence of the dimensionality of our model 
spaces, as 
may be seen in the last entry of Table I, showing the 
percentage of states in the model space that have the 
required quantum numbers.  Furthermore, for the 
$N=6, \Omega=20$ case, there are considerably more $J=2$ states 
than $J=0$, 512 as compared to 137.

%(For $^{22}$Na, an 
%$N=Z=3, \Omega=12$ 
%odd-odd system, we found $T=0$,$J\neq 0$ ground states are favored.)  

{\bf Gaps associated with $J=0$ ground states}:  
In addition to a predominance of $J=0$ ground states, 
such ground states are typically separated by a gap from the 
excited states. A typical case for $N=6, \Omega=20$ 
is shown in Figure 1(a).  Figure 2 
shows the distribution of gaps for $J=0$ ground states.  
The energy is in units of $\bar{v}$, the energy scale used in 
Eqn.~2. 
In these units the centroids of the distributions are  at $\sim {\cal O}( 1)$, 
although 
with a broad width.  For those ground states with $J\neq 0$ the gap is 
much smaller, as shown by an example in Figure 1(b). 
For  $N=6, \Omega=20$,  the average energy gap between a $J=0$ ground state and the first 
excited state is $0.47 \bar{v}$ and 
for $N=4, Z=4, \Omega=12$ it is  $0.79\, \bar{v}$.

{\bf Vibrational/rotational `bands' and yrast structure}: 
In addition to the quantum numbers of the ground state, 
we investigated the evidence for band structure in the low-lying spectra.  
We characterize the low-lying
$J=0,2,4$ yrast states (`yrast' means the lowest state of 
a given angular momentum $J$) with energy $E_J$ by the ratio
$\rho=(E_4 - E_2)/(E_2 - E_0)$.
If an interaction yields a vibrational spectrum, then $\rho=1$, whereas
a rotational spectrum gives $\rho=7/3$. Shown in Fig.~3 is an analysis of
the $J=0,2,4$ spectrum for those samples in our ensemble which had a $J=0$
ground state. All of our cases give broad peaks in the range 
$\rho=0$ to $\sim 1$. 

Note also that some interactions give a $J=0$-$4$-$2$ yrast
character, as indicated by data at $\rho < 0$ in the figure.
Of those samples that exhibit a $J=0$ ground state, approximately 10\%
have a $J=0,2,4$ spin ordering for the three lowest states. 

Although there is no evidence of rotational collectivity among 
the first $J=0,2,4$ states, 
the yrast spectrum extended to high angular momentum shows what is 
called ``noncollective'' rotational behavior in nuclear spectroscopy.
This means that the energies of the yrast states $E_J$ have an 
overall quadratic increase with $J$, but with large fluctuations 
from one $J$ to the next. This is shown in Fig.~4, which displays averaged 
yrast spectra from our ensembles. 
When we fit  $\langle E_J \rangle$ as a function of 
$J(J+1)$ the long-range behavior is roughly linear
with a slope of $0.0539\pm 0.0009 \bar{v}$.

{\bf Phonon collectivity}: 
In  algebraic descriptions of collective behavior one sees far more than 
patterns in the excitation spectra: the low-lying states are  connected
to each other by operators that generate the group representation, 
or at least approximately so, depending on the goodness of the symmetry.  
These operators typically have a large component that is single-particle 
in nature, i.e., expressible as a phonon: 
$ \hat{X}^\dagger = \sum_{\alpha \beta} 
u_{\alpha \beta}  a^\dagger_\alpha a_\beta $
where $a^\dagger, a$ are the usual fermion creation and annihilation 
operators. 
To see whether this collectivity carries over to the RQE we examined 
the transition between the ground $J=0$ states (in the members of the 
ensemble that have such a ground state) and the first excited $J=2$ state. 
For each member of the ensemble we define the phonon  which 
maximally connects these two states by 
$u_{\alpha \beta} = \left \langle J =2  | 
 a^\dagger_\alpha {a}_\beta  | J=0 \right \rangle$.  
We then define the {\it fractional collectivity} $f$ by the ratio of 
the strength of $\hat{X}^\dagger$ to the first excited state to the total 
strength of $\hat{X}^\dagger$ off the ground state, 
\begin{equation}
f = {  \left | \left \langle J  | \hat{X}^\dagger | J=0 \right \rangle 
\right |^2 \over 
\left \langle J=0   | \hat{X} \hat{X}^\dagger | J=0 \right \rangle 
}
\end{equation}
If $f=1$, then the excited state is completely described as a particle-hole 
excitation of the ground state. If $f$ is very small then the two states 
are connected only by many-body operators.   

We studied the fractional collectivity in the $N=4, Z=4, \Omega=12$ 
system, and considered the particle-hole 
phonons between  $J=0$ ground states and the first $J=2$ excited state.
Averaging over our ensemble, we found that 
$\bar{f} = 0.52 \pm 0.27$.  For comparison, the fractional collectivity 
using a realistic nuclear shell-model interaction \cite{Wildenthal}, 
which is known to 
yield strong collectivity among these states, is $0.87$.   For totally 
random states (which we studied by computing the phonon between states 
generated  from 
different interactions) $f \sim 10^{-2}$; this is what one would 
expect from the GOE.  Therefore we find that the 
low-lying states of the RQE can be to a large degree related simply by 
particle-hole excitations.

{\bf Conclusions}
The low-lying spectra of RQE Hamiltonians display markers of 
surprising regularity.  Ground states are predominantly $J=0$ and are 
pushed down relative to the rest of the spectrum, which are two 
of the important characteristics of the BCS pairing Hamiltonian.  
There is also evidence for low-lying vibrational states and noncollective 
rotational when averaged over a large number of yrast states, although 
no evidence for strong rotational collectivity.  Perhaps we should not 
be surprised at these features of the RQE, as it was constructed 
with the Fermi liquid concept of quasiparticles in mind.   The actual 
source of these apparent regularities is not evident, however.  

The numerical studies presented here have barely scratched the surface 
of possible questions that can be addressed with two-body random ensembles.  
Although we have strong evidence for a pair gap, one would 
like to understand analytically whether this persists in the 
large-$N$ limit and its functional dependence on the single-particle space. 
%We considered just two single-particle spaces, 
%$\Omega = 12 = \left\{ {1 \over2}, {3\over 2}, {5\over 2} \right \} $, 
%and $\Omega = 20 =
%\left\{ {1 \over2}, {3\over 2}, {5\over 2}, {7 \over 2}\right \}$, but 
%the average $j$-content of each shell is arbitrary.  If we were to consider 
%spaces of the form $\Omega = 2^n = \left \{  \left ( {1\over 2}\right )^n 
%\right \}$, there would be no orbital angular momentum content in the 
%space, but only the $SU(2)$ symmetry associate with the spin.  
The choice of ensemble will also play a role in the physics.  If the 
ensemble singles out the $J=0$ interaction in the particle-hole channel, 
the Hamiltonian will favor mean-field physics, since the mean field is 
constructed out of the particle-hole density operators.  Thus, we would expect
predictions of mean-field physics, such as the Bethe level density 
formula, to emerge as a limit in this case. Another possibility is to 
emphasize pairing in the particle-hole channel. 
One would then expect to see phonons with more stability than in the RQE, 
and one could explore the more complex group structures that might arise 
(see Ref.~\cite{group}).

We acknowledge discussions with J.~Carlson and W.~E.~Ormand.  CWJ and DJD 
thank the hospitality of the Institute for Nuclear Theory, Seattle, Washington,
where this work was carried out.  
This work is supported by 
Department of Energy grant numbers  DE-FG02-96ER40985, 
DE-FG-06-90ER40561, and 
DE-FG02-96ER40963.
Oak Ridge National Laboratory is managed by Lockheed Martin Energy
Research Corp. for the U.S. Department of Energy under contract number
DE-AC05-96OR22464.

\begin{table}
\caption
{Percentage of ground states of the RQE that 
have $J=0, T=T_z$ for our target nuclides, as compared to 
the percentage of  all states in the model spaces that have 
these quantum numbers. 
%The column labeled 0-2-4 indicates the number of samples which had
%$J=0,2,4$, in that order,  for the first three states.
}
\begin{tabular}{|c|c|c|c|c|}
$N$ & $\Omega $ & nucleus    & $J=0,T=T_z$ & $J=0, T=T_z$\\ 
    &           &           &       g.s.   &    total space \\
\hline
6 & 12&   $^{22}$O      &  $76\%$   &   $9.8\%$     \\
6 & 20 & $^{46}$Ca    &  $75 \%$  & $3.5\%$               \\
$N=4$, $Z=4$ & 12 &  $^{24}$Mg    & $ 66\%$   & $1.1\%$      
\end{tabular}
\end{table}

\begin{figure}
\caption{ `Typical' spectra for 
$N=6, \Omega=20$ ( $^{46}$Ca)  with an RQE Hamiltonian. 
Note the different ground state gaps for ground state $J=0$, $\neq 0$.}
\end{figure}

\begin{figure}
\caption{ Distribution of ground state gaps, defined as the excitation 
energy of the first excited state above a $J=0$ ground state, 
in units of $\bar{v}$ (the energy scale from Eqn.~2).}
\end{figure}

\begin{figure}
\caption{Distribution of $\rho \equiv (E_4 -E_2)/(E_2 -E_0)$ for 
systems with $J=0$ ground state. $\rho = 1$ for vibrational bands and 
$= 7/3$ for rotational bands.}
\end{figure}

\begin{figure}
\caption{Average excitation energy of `yrast' states (lowest state for a 
given $J$) as a function of $J(J+1)$. }
\end{figure}

\bibliographystyle{try}

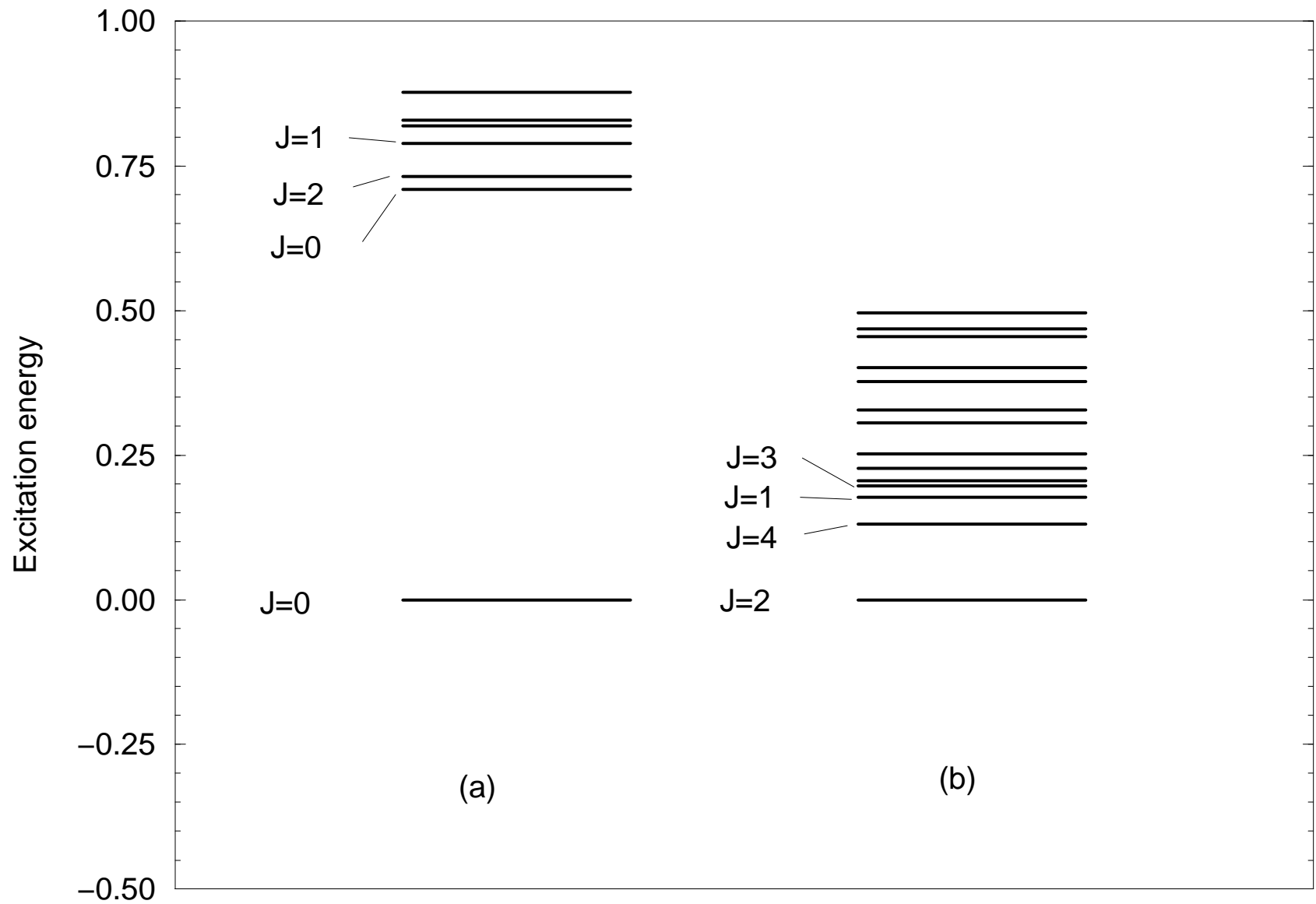
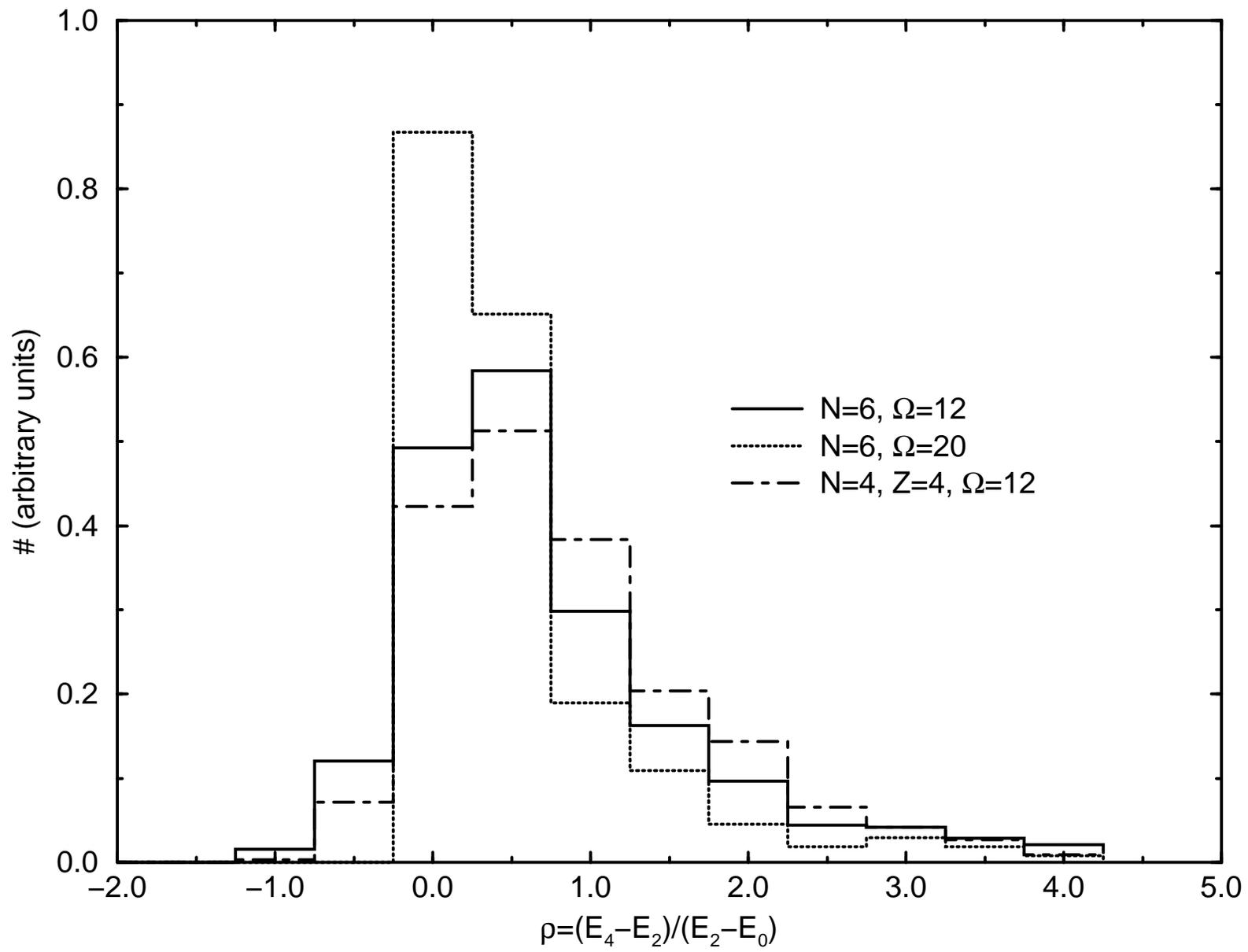
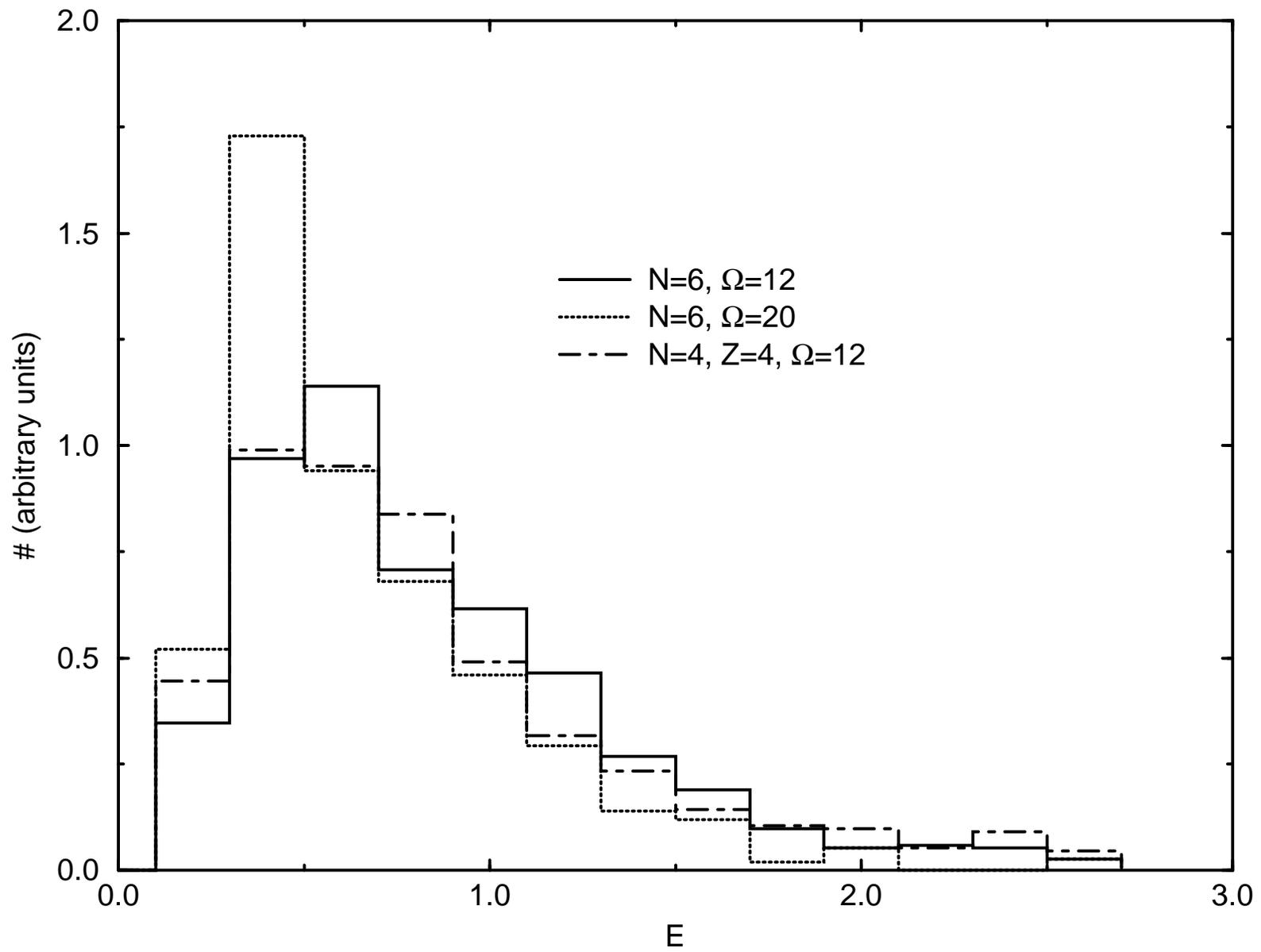
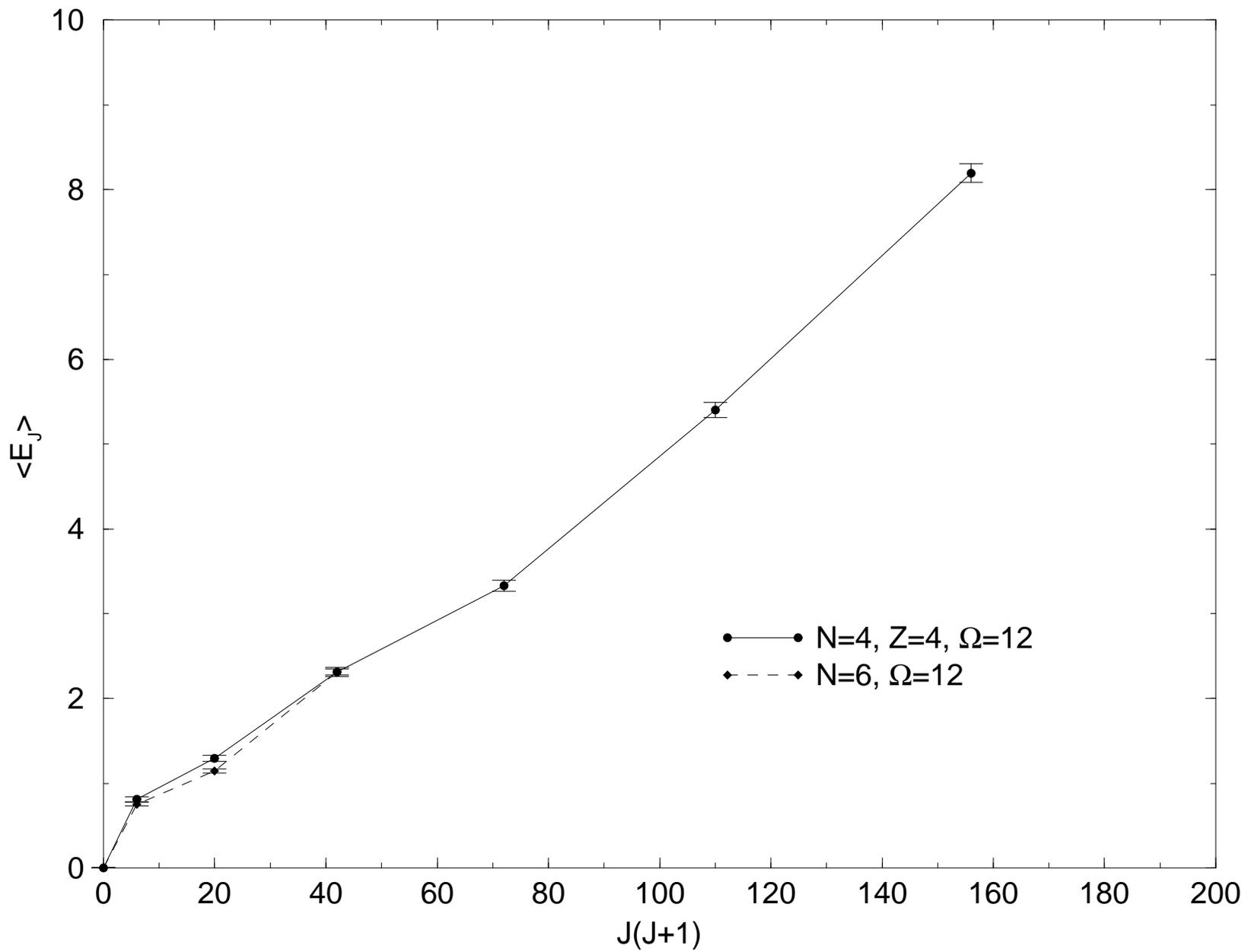
\end{document}